\documentclass[twocolumn,showpacs,aps,prl,10pt]{revtex4}

\usepackage[english]{babel} 	
\usepackage{graphics} 
\usepackage[pdftex]{graphicx}
\usepackage{amssymb} 			
\usepackage{amsmath} 			
\usepackage{booktabs}
\usepackage{tabularx}
\usepackage{rotating}
\usepackage{color}
\usepackage{float}
\usepackage{url}	
\usepackage{bm}
\usepackage{epstopdf}

\begin{document}

\title{Squeezing the Efimov effect}

\author{J.H. Sandoval$^{1,2}$, F.F. Bellotti$^{2}$, M.T. Yamashita$^{1}$,
  T. Frederico$^{3}$, D.V. Fedorov$^{2}$, A.S. Jensen$^{2}$, N.T. Zinner$^{2}$}

\affiliation{$^{1}$Instituto de F\'\i sica Te\'orica, UNESP - Univ
  Estadual Paulista, CEP 01140-070, S\~ao Paulo, SP, Brazil}

\affiliation{$^{2}$Department of Physics and Astronomy, Aarhus
  University, DK-8000 Aarhus C, Denmark}

\affiliation{$^{3}$Instituto Tecnol\'{o}gico de Aeron\'autica,
  12228-900, S\~ao Jos\'e dos Campos, SP, Brazil}

\date{\today }

\begin{abstract}
The quantum mechanical three-body problem is a source of continuing
interest due to its complexity and not least due to the presence of
fascinating solvable cases. The prime example is the Efimov effect
where infinitely many bound states of identical bosons can arise at
the threshold where the two-body problem has zero binding energy. An
important aspect of the Efimov effect is the effect of spatial
dimensionality; it has been observed in three dimensional systems, yet
it is believed to be impossible in two dimensions. Using modern
experimental techniques, it is possible to engineer trap geometry and
thus address the intricate nature of quantum few-body physics as
function of dimensionality. Here we present a
framework for studying the three-body problem as one (continuously)
changes the dimensionality of the system all the way from three,
through two, and down to a single dimension. This is done by
considering the Efimov favorable case of a mass-imbalanced system and
with an external confinement provided by a typical experimental case
with a (deformed) harmonic trap.
\end{abstract}


\maketitle

\paragraph*{Introduction.}
Few-body quantum systems are a theoretical 
and experimental playground for the study of 
the basic structure of quantum mechanics and 
what kind of states are possible in small systems, 
and they also serve as guidance when we want to 
understand many-body problems
\cite{greene-review2017,naidon-review2017,zinner2014,dincao2017}. 
While the 
two-body problem is essentially solvable, at 
least numerically, three interacting quantum 
particles already provide a much more complex, 
and thus interesting, venue for exploration. A 
surprising feature is the Efimov class \cite{efi70}
of infinitely many three-body bound states (trimers) of 
three bosons with resonant short-range two-body interactions in 
three dimensions (3D). This effect has generated tremendous
attention in the last decade due to its observation 
in cold atoms \cite{kraemer2006} 
and lately in Helium trimers \cite{dorner2015}. 
The experimental techniques used to observe such states
are extremely versatile with tunable interactions \cite{chin2010}
geometries \cite{bloch2008,deng2016}, and usage of different atomic 
species \cite{gross2009,knoop2009,zaccanti2009,williams2009,gross2010,lompe2010,nakajima2010,berninger2011,machtey2012,wild2012,knoop2012,roy2013,dyke2013,huang2014}.

A prediction that has not yet been fully explored is the 
fact that the Efimov effect only occurs in 3D and not in 
2D \cite{bruch1979,nielsen1997,brodsky2006,kartavtsev2006,pricoupenko2010,helfrich2011,volosniev2013}. 
(although one may find the so-called super-Efimov effect \cite{nishida2013,volosniev2014,gao2015,efremov2014})
More precisely, 
by performing a well-defined mathematical extension to 
non-integer dimensions, it has been predicted that 
Efimov trimers of identical bosons are only allowed for dimension $d$ in 
the interval $2.3<d<3.8$ \cite{nie01}.
This is a peculiar theoretical prediction that, superficially, appears 
basically inaccessible in actual experiments. 
On the other hand, 
non-integer dimensions play a prominent role in 
for instance high-energy physics \cite{schroder1995} and 
also in low-energy effective field theories \cite{valiente2012}, 
and it would be extremely useful to have a
practical manner in which to study changes in dimensionality 
and how they affect basic quantum few-body physics. 
 
The purpose of the present letter is to investigate how the energies of
the Efimov states in 3D vary as functions of a continuously 
increasing confinement of the spatial dimensions imposed  by
external fields. The infinitely many 3D bound states reduce to a
finite number in 2D, which may be reduced further as 1D is reached.
This provides both qualitative and quantitative answers to the 
question of how much squeezing Efimov trimers can survive, as well
as how trimers will disappear into the continuum. Some recent studies
of Efimov trimers of three identical bosons under confinement have been reported \cite{lev14,yam15}, 
as well as earlier work on fermions in quasi-2D \cite{levinsen2009} and 
mixed-dimensional confinement \cite{nishida2008}.
However, no previous 
study has been able to provide continuous dimensional squeezing from 
3D to 2D, and all the way down to 1D with non-identical particles. 
Furthermore, the formalism we present can be applied to any confinement
geometry in principle. Here we focus on the most widely applied 
experimental situation with a deformed harmonic confinement, and on 
mass asymmetric systems which are a current focus of three-body 
physics \cite{barontini2009,bloom2013,pires2014,tung2014,maier2015,ulmanis2016,wacker2016,johansen2016}.

\paragraph*{Method.}
We consider an $AAB$ system with two identical (bosonic) $A$ 
particles of mass $m_A$ and a $B$ of mass $m_B$. The reduced mass
is defined by $\mu=m_A m_B/(m_A+m_B)$. In order to reduce the number
of parameters, we assume that the $A$ particles are not interacting,
while the $AB$ subsystem has a short-range interaction that we model
by a Gaussian potential, $-S_0 \exp(-\bm{r}^{2}_{AB}/r_{0}^2)$, where
$\bm{r}_{AB}$ is the relative coordinate of the $AB$ system. The
non-interacting nature of the $AA$ system is a matter of convenience
and not essential as our formalism applies to general systems (see
\cite{supmat} for details).  The interaction range, $r_0$, is kept
small while the strength, $S_0>0$, is tuned so that it reproduces a
fixed 3D (vacuum) scattering length, $a_\textrm{3D}$, in the region
close to the resonance at $2\mu b^2 S_0/ \hbar^2=2.68$ where
$|a_\textrm{3D}/b|=\infty$. For concreteness, we focus on the case
where $a_\textrm{3D}>0$ so that a two-body bound state with small
binding energy, $E_{2}^{\textrm{3D}}= \hbar^2/(2\mu a_\textrm{3D}^2)$,
exists. In order to squeeze the system, we assume the same external
one-body harmonic oscillator potential on each particle along two
directions, $\tfrac{1}{2}m(\omega_{x}^{2}x^{2}+\omega_{y}^{2}y^{2})$,
where $m$, $x$ and $y$ are mass and single-particle Cartesian
coordinates of particles $A$ or $B$.  For simplicity we
use identical external confinement on each particle as this 
decouples the center-of-mass motion
\cite{supmat}. We expect the physics to remain qualitatively the 
same with unequal trapping. Defining $b_x=\sqrt{\hbar/\mu \omega_x}$ and
$b_y=\sqrt{\hbar/\mu \omega_y}$, we squeeze the system starting from
large values of $b_x$ or $b_y$ and decreasing these towards $b_x\to 0$
or $b_y\to 0$.

In order to solve the three-body problem we use a momentum-space
approach and the integral Faddeev equations
\cite{adh95a,adh95b,fre11}.  These equations are modified to allow for
squeezing by imposing periodic boundary conditions along one or
several directions, effectively compactifying those dimensions on a
ring of radius $R_{x/y}$. This implies that the momenta along the
compact directions are discrete. In the limit where $R_{x/y}\to 0$,
the gap in the spectrum along a compact dimension goes to infinity,
which eliminates motion in that direction, whereas in the limit
$R_{x/y}\to\infty$, the gap vanishes and we recover the usual 
continuous spatial $x/y$ dimension. 
The results presented in this letter show
that this formalism is capable of addressing the full crossover
between different (integer) dimensions for general three-body systems
of any mass.

The concrete implementation of our compactified Faddeev equations uses
effective zero-range interactions. However, as is well-known from
previous three-body Efimov studies \cite{efi70}, the decisive
parameter(s) are the two-body binding energies between pairs of
particles, which are typically parameterized by $a_{3D}$. 
In our setup, we have $AB$ interactions with two-body
energy $E_{2}^{\textrm{3D}}$.  It is important to stress that our
input is the two-body energy calculated in a fully 3D setup that 
{\it includes} the external confinement. This is done by calculating
$E_{2}^{\textrm{3D}}$ using a correlated Gaussian
numerical technique \cite{mitroy2013} with fixed $a_\textrm{3D}$ while varying the
trap by decreasing for instance $b_y$. We then relate $b_y$ and
$R_y$ by demanding that
$E_{2}^{\textrm{3D}}(b_y)=E_{2}^{\textrm{3D}}(R_y)$, where the latter
is calculated with a compact dimension (see \cite{supmat} for
details).  Numerically, we find the remarkably simple result $b_y
\approx 2\pi R_y$, and clearly see that $b_y\to 0$ will correspond to
the 2D limit as expected.  Further squeezing from 2D down to 1D is
accomplished by starting from a 2D version of the Faddeev equations
\cite{bel11,bel12,bel13b} and is otherwise analogous \cite{supmat}.  This
method can be extended to other kinds of confinement through the two-body
subsystems.

\begin{figure}[ht!]
\includegraphics[width=1\linewidth]{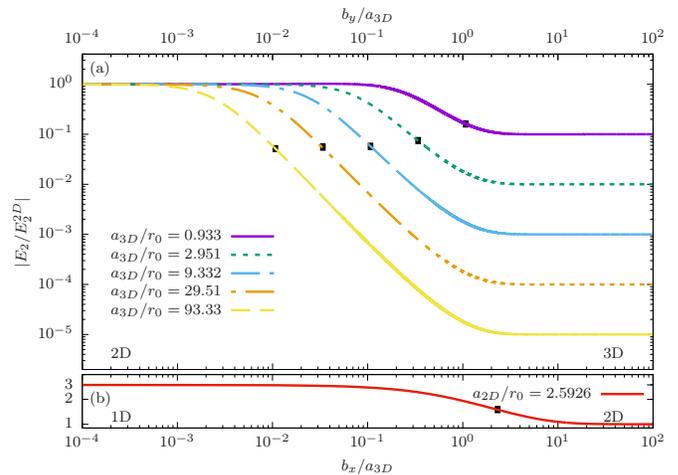}
\caption{
(a) The two-body energies normalized to the 2D limit ($E_{2}^{2D}$) as functions
of $b_{y}/a_{3D}$ for different $a_{3D}/r_0$. The black points indicate 
where $b_{y} = r_0$. (b) The corresponding 2D to 1D transition as 
function of $b_x/a_{3D}$. The length scale $a_{2D}$ is defined through
$|E_{2}^{2D}|=4e^{-2\gamma}\hbar^2(\mu a_{2D}^{2})^{-2}$, where $\gamma$ is
Euler's constant.
}
\label{fig1a}
\end{figure}

\paragraph*{Two-body properties.}
We first consider the $AB$ two-body subsystem. 
The energy as function of $b_y/a_\textrm{3D}$ for fixed
$a_\textrm{3D}$ is shown in Fig.~\ref{fig1a}(a) where we have normalized
to the energy in the 2D limit ($b_y\to 0$).  In
Fig.~\ref{fig1a}(a) we see an evolution from the 3D limit (far right
side) with energies that remain constant until around the point where
$b_y\sim r_0$. This is when the external confinement starts to be felt
strongly by the particles and the energy moves quite fast towards the
2D limiting value. It is interesting to note that the energy at which
$b_y=r_0$ (marked by black points in Fig.~\ref{fig1a}) is almost the
same, $E_2/E_2^\textrm{2D}(b_y=r_0)\sim 0.05$, independent
of $a_\textrm{3D}$ for $a_\textrm{3D}/r_0\gg 1$. 
The evolution from
2D to 1D is shown in Fig.~\ref{fig1a}(b) and confirms our expectation 
that further binding occurs as we approach the 1D limit.

\paragraph*{Spectral flow from 3D to 2D.}
We now proceed to discuss Efimov trimer states as we continuously
squeeze along one direction, i.e. as $b_y$ decreases. The mass ratio
is taken to be $m_B/m_A=6/133$ \cite{note-on-dim} and is relevant for current studies of
trimers in $^6$Li-$^{133}$Cs mixtures
\cite{pires2014,tung2014,ulmanis2016,johansen2016}. This gives a relatively small Efimov
scaling factor $e^{\pi/s}=4.788$ \cite{jen03,yam13} so that many
Efimov trimers can be expected. We choose a large $a_\textrm{3D}/r_0
\simeq 10^5$ to perform our calculations.

\begin{figure}[ht!]
\includegraphics[width=1\linewidth]{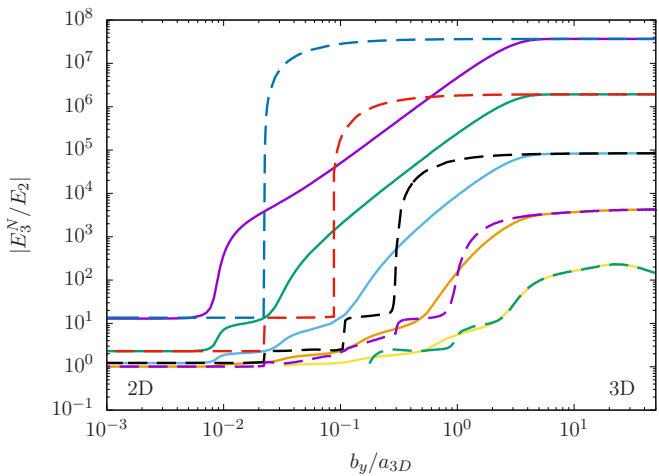}
\caption{Trimer energies plotted in units of the 
two-body energy for $m_B/m_A = 6/133$ 
as functions of $b_{y}/a_\textrm{3D}$. For the solid lines
the two-body energy varies with $b_y$ while for 
the dashed lines it is kept constant (see text for discussion).
Solid and dashed lines have different colors for visibility.
}
\label{fig3a}
\end{figure}

The three-body energies of the $N$th trimer,  $ E_{3}^{N}$, 
relative to the two-body energy are shown in 
Fig.~\ref{fig3a} as function of $b_y/a_\textrm{3D}$. 
In the 3D limit to the far right, we are
able to numerically resolve five Efimov states which scale 
in energy with $e^{2\pi/s}$ as expected. In the strict 
2D limit on the far left of Fig.~\ref{fig3a}, we find that 
four states survive as expected \cite{bel11}. The behavior 
in between these two integer limits is intriguing and depends
sensitively on how we treat the two-body energy. 

The dashed lines in Fig.~\ref{fig3a} show the results obtained
when assuming that the two-body energy does not vary with $b_y$
and is set by the 3D value, $E_2=E_{2}^\textrm{3D}(b_y\to \infty)$.
As $b_y$
decreases we see a number of systematically occurring abrupt drops in
$E_{3}^{N}$. Each drop is from an initial value down to one of the energies
that the system is destined to reach in 2D where the Efimov effect is gone.

Specifically, as we decrease $b_y$ (going from right to left in Fig.~\ref{fig3a})
the state that is weakest bound in the 3D limit 
first decreases its energy to a value corresponding to the 
strongest bound state in the 2D limit. It then has roughly constant
energy until the next level decreases its energy and demands the
position in the spectrum, and pushed the state down to an energy
around that of the first excited state in the 2D limit.
These
processes are repeated until the four 2D positions
are reached and the remaining three-body state has
disappeared into the continuum (a single state in our case). 
They are reminiscent of the so-called Zeldovich rearrangement
\cite{zeldovich1960}, in which the short-range interactions compete with 
the long-range influence of the confinement.

It is important to notice that before these abrupt changes 
of the energies, the Efimov scaling among the states is 
intact. Thus, we have a quantitative measure of how much 
squeezing different Efimov states can survive. A rough estimate
of the jumps can be inferred by considering the Efimov 
attractive inverse square potential which extends to 
around $a_\textrm{3D}$ \cite{jensen2004,braaten2006}, and therefore the 
radial extent of the least bound state is roughly $a_\textrm{3D}$.
In turn, the first spectral jump is expected around $b_y\sim a_\textrm{3D}$, 
since here the state becomes strongly influenced by the trap \cite{portegies2011}.
Subsequent jumps now follow an Efimov scaling law and occur when
$b_y\sim a_\textrm{3D}/e^{N\pi/s}$.

Keeping a constant $E_2$ value is presumably experimentally challenging
as it requires tuning of interactions to compensate for the effects of
the confinement on $E_2$. We therefore now study the case where this
is not done so that we now have a varying $E_{2}(b_y)$.  This changes
the flow as seen in Fig.~\ref{fig3a}.  The decrease of energies will
start for larger values of $b_y$ and have a considerably smoother
behavior.  Remarkably, we see that the energy curves are roughly
parallel on a double-log scale, thus showing that even in this case we
have signatures of Efimov scaling prominently featured.  We stress
that, even though the abrupt changes found for a constant $E_2$ are
now smoother, we still clearly see the rearrangements discussed above,
and these features could be a very clear experimental signature to
confirm the present predictions.

\begin{figure}[ht!]
\includegraphics[width=1\linewidth]{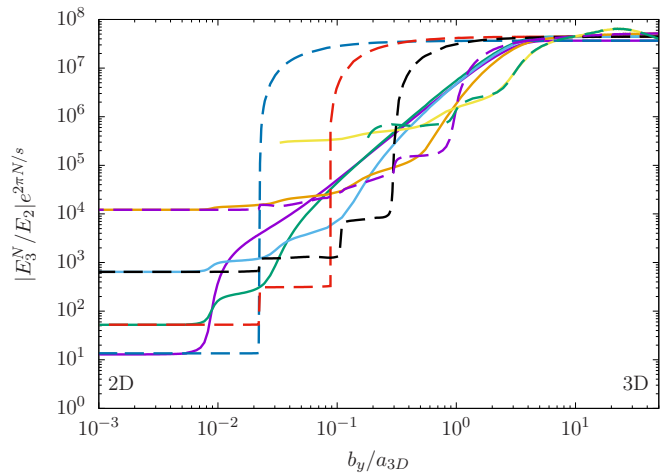}
\caption{
Energies as in Fig.~\ref{fig3a} but now multiplied by
the scaling factor $\exp(2\pi N/s) = 22.92^N$ where $N=0,1,2,3,4$
for ground and excited states.}
\label{fig3c}
\end{figure}

In order to investigate the Efimov scaling as function of the
squeezing, we now multiply the three-body energies by $e^{2\pi N/s}$
for the $N$th Efimov state in the energies.  The results are shown in
Fig.~\ref{fig3c}.  For the case of constant $E_2$ the results are very
similar to Fig.~\ref{fig3a}, while those with varying $E_2(b_y)$ now
more clearly shows a tendency to collapse onto a single curve over an
extended region. This region is limited by the necessity for the
states to match up with their 2D limiting values, and they each leave
the common curve due to rearrangements one at a time starting from the
weaker bound state.  We can infer from the dashed lines in
Fig.~\ref{fig3c} that a scaling of $e^{2\pi N/s}$ on $b_y$ would tend
to also collapse the case of constant $E_2$ onto a single curve. This
is not needed when $E_2$ varies. The intriguing conclusion appears to be
that the two-body subsystem ($E_2(b_y)$) already contains the information on the scaling.

\begin{figure}[t]
\includegraphics[width=1\linewidth]{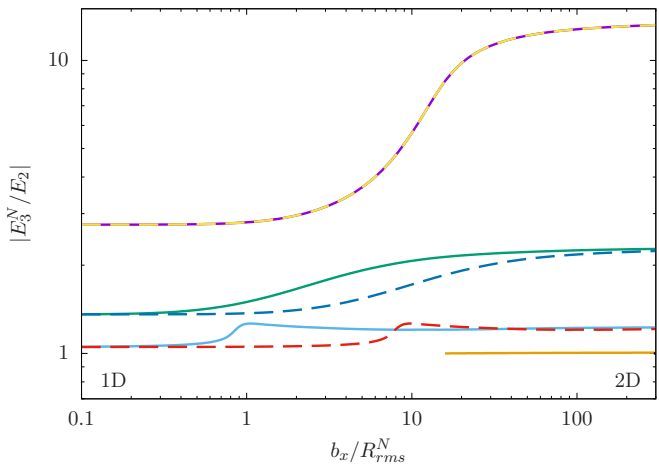}
\caption{
The ratio of three- and two-body energies for 
$m_B/m_A = 6/133$ as functions of $b_x /R_{rms}^N$. 
The
dashed curves for all states are with $R_{rms}=R_{rms}^0 = 0.4742 b =
0.2101\hbar/\sqrt{m_AE_{AB}(2D)}$ (see text for discussion).
Solid and dashed lines differ in color for visibility. 
}
\label{fig3b}
\end{figure}

\paragraph*{Squeezing down to 1D.}
Starting from the 2D limit results shown in Fig.~\ref{fig3a}, we may
consider what happens as we further squeeze the system down to 1D by
increasing the harmonic confinement along the $x$-direction.  
Technically, we start from 2D Faddeev
equations and proceed as before (see \cite{supmat} for details).
The results of this are shown in Fig.~\ref{fig3b}.  We notice similar
behavior with plateaus in 2D and 1D limits connected by intermediate
transitions where the energy changes rapidly. Notice that for the mass
ratio used, the 1D limit only holds three bound states, and one state
goes to the continuum during the dimensional reduction.

As Efimov scaling does not extend to these low dimensions, the natural
quantities to analyze the system are slightly different. As
was recently discussed in Ref.~\cite{san16}, the root-mean-square radius of
the $N$th $2D$ three-body state, $R_{rms}^N$, is proportional to the
inverse square root of $E_{2}^\textrm{2D}$, with a proportionality
factor that depends on the state index $N$ and the mass ratio. Since
the three-body equations depend only on the quantity $E_{2}(b_x)$
\cite{supmat}, the three-body energy must be a function of
$E_{2}(b_x)$. In turn, the three-body to two-body energy ratio will depend
only on $b_x/R_{rms}^N$ for the $N$th state. The transition from 2D to
1D can therefore be studied in a universal manner by using this
variable as done in Fig.~\ref{fig3b}.  For comparison, we plot the
energies as function of $b_x/R_{rms}^0$ with dashed lines in
Fig.~\ref{fig3b}, in order to follow each state for the same value of
$b_x$.

The bound state behavior under squeezing from 2D to 1D is 
clearly different from the case of 3D to 2D. In particular, 
we see in Fig.~\ref{fig3b} that all the three states that 
survive to the 1D limit start to feel the squeezing already 
for relatively large traps. If we focus on the 
dashed lines, we see that the center of the drop is around 
$b_x/R_{rms}^0\sim 10$ for all of the states, indicating 
that we have a synchronized pattern of rearrangements in 
contrast to the hierarchical pattern seen in Fig.~\ref{fig3a} 
and Fig.~\ref{fig3c}. In Fig.~\ref{fig3b}, the stronger 
bound state gets pushed to its 1D limit and forces the 
other states to follow suit. However, it is still very 
clear that there is a sizable effect of the squeezing 
that should be observable.

\paragraph*{Experimental implications and outlook.}  
Observing the influence of squeezing on the Efimov effect and the
spectral flows that this generates should be possible with the
experimental techniques that have hitherto been used to probe
three-body physics with cold atoms. A much used tool is recombination
rate studies where three-body states are identified by peaks and
interference minima in the rate.  In the case of squeezing from 3D to
2D, we have a finite $b_y$. We may now vary $a_\textrm{3D}$ while
keeping $b_y$ fixed which will scan from right to left in Fig.~\ref{fig3a},
and would expect to see a feature in the recombination rate around the
point where the least bound state enters the continuum. Here we use
that the flow depends solely on $b_y/a_\textrm{3D}$, but we note that
the number of bound states to work with depends on how large initial
value of $a_\textrm{3D}$ one can access in a concrete
experiment. Similarly, if we consider an experiment where $b_y$ is
tuned independently of $a_\textrm{3D}$, then we may take a fixed ratio
$b_y/a_\textrm{3D}$ and vary $a_\textrm{3D}$ which will cause bound
states to cross into the continuum. Doing so for several different
values of $b_y/a_\textrm{3D}$ would allow verification of our
predictions. The same method can be applied in the case where we go
from 2D to 1D.  An alternative to recombination rate measurement is to
use radio frequency association \cite{lompe2010,nakajima2010} to access the binding
energies themselves. This is more difficult but also yields more
information. In this case one should be able to observe the spectrum
at several points by varying $b_y$ and/or $a_\textrm{3D}$ to see the
flow of the states.

\paragraph*{Outlook.}  
In the present work we have focused our attention on a simple setup 
in order to best illustrate the effects of squeezing on the energies of Efimov 
trimers. Our formalism can be used to 
discuss other quantities such as radial extension of states, 
momentum distributions etc. We have also chosen a particular 
mass ratio that corresponds to recent experiments, but simplified
our discussion by neglecting interactions between the two heavy
particles in the trimer. While we do expect quantitative changes
when including this interaction, the qualitative behavior should 
be the same. Likewise, we expect the same behavior as discussed
here in the case where $a_{3D}$ is large but with negative sign.  
A cylindrical confinement may also be accommodated by a simple 
modification of our formalism and this will allow squeezing 
of the system along two directions ($b_x=b_y\to 0$). Initial 
investigations indicate that a direct transition from 3D to 1D
yields similar results to those presented above.

\paragraph*{Acknowledgments.} 
The authors would like to thank A.~R. Rocha, S.~J.~J.~M.~F. Kokkelmans, 
and J. Levinsen for feedback on the results and the manuscript.
This work was partly supported by funds 
provided by the Brazilian agencies Funda\c{c}\~{a}o de Amparo \`{a} 
Pesquisa do Estado de S\~{a}o Paulo - FAPESP grant no. 2016/01816-2(MTY), 
Conselho Nacional de Desenvolvimento Cient\'{i}fico e Tecnol\'{o}gico - CNPq 
grant no. 302075/2016-0(MTY), Coordena\c{c}\~{a}o de Aperfei\c{c}oamento 
de Pessoal de N\'{i}vel Superior - CAPES no. 88881.030363/2013-01(MTY), 
and by the Danish Agency for Science, Technology, and Innovation.

\newpage
\widetext
\begin{center}
\textbf{\large Supplemental Material for ``Squeezing the Efimov Effect''}
\end{center}

\section{Squeezed dimer}
In this section we present the equations that are used to obtain the dimer energy 
as we squeeze along one (3D$\rightarrow$2D) 
or along two (2D$\rightarrow$1D) spatial dimensions. We will be using units where $\hbar=1$ throughout the 
discussion in this supplementary material.

\subsection{Transition from 3D$\rightarrow$2D}
In our model we will assume periodic boundary conditions along 
one direction (chosen to be the $y$-axis). Then, the relative momenta along the plane 
are given by $\vec p_\perp=(p_x,p_z)$ and
\begin{eqnarray}
p_y=\frac{n}{R_y} \ , \label{eq1}
\end{eqnarray}
with $n=0,\pm1,\pm2,\dots$. The length of the squeezed dimension corresponds to a 
radius, $R_y$, that interpolates between the 2D limit for $R_y\to 0$ and the 3D limit 
for $R_y\to \infty$. As discussed in the main text, the choice of a periodic dimension 
is not essential for our study, as we may map the physics of other types of external 
confinement onto the system with periodic boundary conditions. In the present case
we consider the case of a harmonic oscillator confinement that we map onto the 
periodic setup.

First, we consider the case where we have zero-range (ZR) interactions. 
In general, the dimer energy with zero-range interactions $E_{2}^{ZR}$ is a function of $R_y$. A natural fixed point of 
the dimer energy is the 3D limit where the shallow zero-range dimer energy
around for instance a Feshbach resonance is experimentally measurable. We denote this dimer
energy of a 3D setup (no squeeze) by $E_2^{3D}$. 
This implies that the two-body $T$-operator in the limit 
$R_y\to\infty$ has to recover a pole exactly at $E_{2}^{3D}$. Thus, for the 
zero-range potential we must solve \cite{ziegelmann}
\begin{eqnarray}
\int d^3p\frac{1}{E_2^{3D}-\frac{p^2}{2M}} -
\frac{1}{R_y}\sum_n\int d^2p_\perp\frac{1}
{E_{2}^{ZR}-\frac{p_\perp^2}{2M}-\frac{n^2}{2MR_y^2}}
=0 \ , \label{rtau1}
\end{eqnarray}
where $M$ is the reduced mass of the dimer. The above equation can be solved analytically giving:
\begin{equation}
\sqrt{-ME_{2}^{ZR}}
= \frac{1}{\pi R_y}\sinh^{-1} \frac{e ^{\pi\, R_y/a_{3D}}}{2} \ ,
\label{rtau7}
\end{equation}
where $a_{3D}=\sqrt{-E_{2}^{3D}}$ is the two-body scattering length. The explicit form of (\ref{rtau7}) reads:
\begin{equation}
E_{2}^{ZR}=- \frac{a_{3D}^{2}}{(\pi R_y)^2}\ln^2\left(\frac{e ^{\pi\, R_y/a_{3D}}}{2}+\sqrt{\frac{e ^{2\pi\, R_y/a_{3D}}}{4}+1}\right) \ ,
\label{e2expl}
\end{equation}
and for $R_y\to 0$ one has that, for a zero-range potential, the dimer energy changes as: 
\begin{equation} \label{e2zr}
E^{ZR}_2\sim -(\pi MR_y)^{-2}\,  (\sinh^{-1} \frac{1}{2})^2  = 0.02346227\, \, (MR_y)^{-2}. 
\end{equation}
This result should not be valid for a finite-range potential, as in this case we expect a 
finite dimer energy when the system is confined in two dimensions ($R_y\to0$).  

The argument above shows that the route toward $R_y\to 0$ depends on the form of the two-body potential. 
In order to regularize $E^{ZR}_2$ for $R_y\to0$, we assume a simple fitting formula for 
the dimer energy as a function of $R_y$. This formula will have two parameters constrained 
to the dimer binding energies calculated numerically at the 2D and 3D limits as we will now 
discuss.

To calibrate the zero-range model, we use the numerically highly robust stochastic 
variational method to calculate the dimer binding energies in the presence of 
a harmonic trap which is then squeezed along one direction. The zero-range 
interaction is modeled by a Gaussian two-body potential. 
Thus, we solve the following eigenvalue equation
\begin{equation}
H|\Psi\rangle=\left(\frac{p_{AB}^{2}}{2M}
+V(r_{AB})+\frac{M}{2}\left(\omega_{x}^{2} x_{AB}^{2}+\omega_{y}^{2} y_{AB}^{2}\right)\right)|\Psi\rangle\,=\,
e_2|\Psi\rangle \, , \label{e2bw2} 
\end{equation}
where $V(r)=S_0e^{-r_{AB}^{2}/r_{0}^{2}}$ is the two-body interaction at (relative) distance $r_{AB}$ with
strength $S_0$ and range $r_0$. When we squeeze from 3D to 2D, we take $\omega_x=0$ and increase $\omega_y$. 
Note that $x_{AB}$ and $y_{AB}$ are the Cartesian components of the relative distance between the 
two particles, $r_{AB}$. The center of mass part of the trap decouples from the problem and can be 
ignored in our case where we are only interested in the intrinsic internal dynamics of dimer
and trimer states. The energy 
$e_2=\frac{\langle\Psi|H|\Psi\rangle}{\langle\Psi|\Psi\rangle}$ is calculated from a correlated 
Gaussian basis used to expand the wave function \cite{suzuki}.

Equation~\eqref{e2bw2} is now used to define $E_2(\omega_y)= e_2-\frac{\hbar\omega_y}{2}$. The
subtraction of the zero-point contribution is important as one would otherwise get
a divergent contribution that would reflect only the increasing trap energy and not 
the intrinsic behavior of the dimer. We find that the dimer energy is accurately 
described by the form 
\begin{equation}
E_2(b_y)=- \frac{4a_{3D}^{2}}{\alpha+\beta b_{y}^2}\ln^2
\left(\frac{e ^{b_y/2a_{3D}}}{2}+\sqrt{\frac{e ^{b_y/a_{3D}}}{4}+1}\right),
\label{zrreg}
\end{equation}
where we have defined the oscillator length $b_y=\sqrt{1/M\omega_y}$. This form 
is of course inspired by the zero-range dimer energy above, Eq.~\eqref{e2expl}.
In order to fix the parameters, $\alpha$ and $\beta$, we may use the limiting 
expressions $E_2(b_\omega\to0)\equiv E_2^{2D}$ and $E_2(b_\omega\to\infty)\equiv E_2^{3D}$, 
which gives 
\begin{equation}
\alpha\equiv-\frac{4a_{3D}^{2}}{E_2^{2D}}\ln^2\left(\frac{1+\sqrt{5}}{2}\right) \,{\rm and}\, 
\beta\equiv-\frac{1}{E_2^{3D}}.
\end{equation}
By comparison between Eq.~\eqref{e2expl} and Eq.~\eqref{zrreg}, we may now infer that 
the mapping between our setup with periodic boundaries to that of the harmonic 
trap is obtained by identifying $2\pi R_y=b_y$. Numerically, we find that this 
relationship is extremely accurate. 

The procedure above may be performed for other confinement potentials with 
little extra complication as the stochastic variational method is highly flexible \cite{mitroy2013}
and can provide the necessary dimer energies that we need to calibrate our setup with 
zero-range interactions and periodic boundaries. The precise mapping relation 
between $R_y$ and the length parameters of other confining potentials may of course 
differ from that presented here.

\subsection{Transition 2D$\rightarrow$1D}
In this section we squeeze one of the two remaining directions of the last subsection 
confining the dimer in 1D. This corresponds to now increasing $\omega_x$. 
We repeat essentially the same steps to obtain the binding energy of the 
dimer as a function of $R_x$. Eq.~\eqref{rtau1} is changed so that it describes the $2D\to1D$ 
transition. In this case it has the form
\begin{equation}
\int  
\frac{d^2p}{E_2^{2D}-\frac{p^2}{2M}}- 
\frac1R_x\sum_n\int\frac{dp}{\bar E_2-\frac{p^2}{2M}-\frac{n^2}{2MR_x^2}}=0,
\label{rtauab1da}
\end{equation}
where the two-body binding energy is written with a bar 
$\bar E_2$ and now depends implicitly on $R_x$.
After performing the sum over the discrete modes and the integration 
over one of the momenta, we get the following transcendental equation 
for the two-body binding energy as:
\begin{equation}
\ln\left(\frac{\bar E_2}{E_2^{2D}}\right)
 =2\int^\infty_{0}dp\frac{\coth (\pi\, R_x \sqrt{-\bar E_2+p^2/2M})-1}{\sqrt{-\bar E_2+p^2/2M}}
\ . \label{pole2d1b}
\end{equation}  

In the limit $R_x\to\infty$, Eq.~\eqref{pole2d1b} reproduces the two-body energy in $2D$. However, 
it diverges in the limit $R_x\to0$ and needs to be regularized. This is 
done by replacing $R_x^2\to R_x^2+R_0^2$, in which $R_0$ is an adjustable 
parameter that allows us to obtain $\bar E_2(R_x=0)=E_2^{1D}$, where $E_2^{1D}$ is the two-body energy in $1D$
which is calculated via the the stochastic variational method using a Gaussian potential 
just as we have done in the previous section. The mapping is again found to be 
$2\pi R_x \approx b_x$, where $b_x=\sqrt{1/M\omega_x}$.

\section{Squeezed trimer}
The trimer we now consider is an $AAB$ system with two identical $A$ particles 
of bosonic kind, and a third particle $B$ that may have a different mass.
In what follows we detail the integral equations for the bound state, 
in which we introduce a compact dimension through a periodic boundary condition quantizing the 
relative momentum of the third particle with respect to the interacting pair. Furthermore 
 the two-body amplitudes for a given squeezing situation are defined such that the two-body bound 
state energies come from Eqs.~\eqref{zrreg} and \eqref{pole2d1b} for the transition 3D$\to$2D and
2D$\to$1D, respectively. 

\subsection{Transition 3D$\rightarrow$2D}
To describe the trimer, 
we will use relative Jacobi coordinates, where 
$\vec p$ represents the relative momentum of a given pair of 
particles in the three-body system and $\vec q$ the momentum 
of the remaining particle with respect to center of mass of said pair. We are 
interested in the universal limit 
where the ranges of all two-body interactions can be neglected. This means
that we consider zero-range interaction as in the dimer case above. 
Zero-range interactions present 
a singularity which is resolved by a subtraction in the kernel with 
the introduction of a scale, $\mu^2$ \cite{adhikari}. For simplicity we will use units 
where $\hbar=m_A=1$ from now on and  introduce the mass number ${\cal A}=m_B/m_A$. 
We will denote the three-body trimer binding energy by $E_3$ in the following. 

The coupled and subtracted integral equations for the 
spectator functions, $f_{AA}$ and $f_{AB}$, of the trimer system can be 
written down in the case 
where one direction of the relative 
momenta, $\vec q$ and $\vec p$, are quantized in the manner 
outlined in Eq.~\eqref{eq1}. They
are given by 
\begin{eqnarray}
\nonumber
&&f_{AA}(\tilde q)=-2\,\tau_{AA;R_y}\left(E_3-\frac{{\cal A}+2}{4{\cal A}} \tilde q^2\right) 
\sum_m\int\frac{d^2p_\perp}{R_y} 
\left[G^{(1)}_{0R_y}(\tilde q,\tilde p;E_3)-G^{(1)}_{0R_y}(\tilde q,\tilde p;-\mu^2)\right] f_{AB}(\tilde p)\\ \nonumber
&&f_{AB}(\tilde q)=-\tau_{AB;R_y}\left(E_3-\frac{{\cal A}+2}{2({\cal A}+1)}  \tilde q^2 \right) 
\sum_m\int \frac{d^2p_\perp}{R_y} \left\{
\left[G^{(1)}_{0R_y}(\tilde p,\tilde q;E_3)-G^{(1)}_{0R_y}(\tilde p,\tilde q;-\mu^2)\right] f_{AA}(\tilde p)
\right. \\ 
&&\left.+ \left[G^{(2)}_{0R_y}(\tilde q,\tilde p;E_3)-G^{(2)}_{0R_y}(\tilde q,\tilde p;-\mu^2)\right] f_{AB}(\tilde p)\right\}
\label{zre5a}
\end{eqnarray}
where  $\tilde q\equiv (\vec q_\perp,n)$, $\tilde p\equiv (\vec p_\perp,m)$ and 
\begin{equation}
\nonumber
\tilde q^2=q_\perp^2+{n^2\over R_y^2}\,\, ,\,\, \tilde p^2=p_\perp^2+{m^2\over R_y^2} \,\, {\rm and} \,\,  
\tilde q\cdot \tilde p=\vec q_\perp\cdot \vec p_\perp +\frac{n\;m}{R_y^2}
\end{equation}
The resolvents are defined by:
\begin{equation}
\left[G^{(1)}_{0R_y}(\tilde q,\tilde p;E)\right]^{-1}=
E-\tilde p^2-\tilde q\cdot \tilde p-\frac{{\cal A}+1}{2{\cal A}}\tilde q^2 
 \,\,\,\,\, , \,\,\,
\left[G^{(2)}_{0R_y}(\tilde q,\tilde p;E)\right]^{-1}= E
-\frac{\tilde q\cdot \tilde p}{{\cal A}}-\frac{{\cal A}+1}{2{\cal A}}(\tilde q^2+\tilde p^2)
\label{greenq2d}
\end{equation}
The two-body amplitudes for finite $R_y$ are given by
\begin{equation}
R_x \,\tau^{-1}_{A\beta;R_y}(E)
 =2\,m_{A\beta}\,\left\{\sum_n\int  
\frac{d^2p_\perp}{ \tilde E-
p_\perp^2-\frac{n^2}{R_y^2}}
-\sum\int  
\frac{d^2p_\perp}{\tilde E_{A\beta}-
p_\perp^2-\frac{n^2}{R_y^2}} \right\}
\ , \label{rtauab}
\end{equation}
with $\beta\equiv A$ or $ B$, $\tilde E=2\,m_{A\beta} E$ ($E<0$) and $\tilde E_{A\beta}=2\,m_{A\beta} E_{A\beta}$ 
and we chose the bound-state pole at $E_{A\beta}$ for each $R_y$. The reduced mass is $m_{A\beta}=m_A\, m_\beta/(m_A+ m_\beta)$.
Performing the analytical integration over $\vec p_\perp$ and performing the sum, we get that
\begin{eqnarray}
\tau_{A\beta;R_y}(E)=R_y\left[
4\pi\,m_{A\beta}\ln\left({\sinh\pi\sqrt{-2\,m_{A\beta}E}\,R_y\over\sinh\pi\sqrt{-2\,m_{A\beta}E_{A\beta}}\,R_y}\right)\right]^{-1}
\label{tauq3d}\end{eqnarray}

In  the limit of $R_y\to\infty$  the two-body amplitudes for $AA$ and $AB$ 
reduces to the known 3D expressions
\begin{equation}
\tau_{A B;R_y\to\infty}(E)\equiv
\frac{1}{2\pi^2}\left(\frac{{\cal A}+1}{2{\cal A}}\right)^{3/2}
\left[\sqrt{-E} -
\sqrt{-E_{AB}}\right]^{-1}  \,\,\, ,\,\,\,\,
\tau_{A A;R_y\to\infty}(E)\equiv \frac{1}{2\pi^2}
\left[\sqrt{-E} - \sqrt{-E_{AA}}
\right]^{-1}  \label{taua}
\end{equation}
In the 3D limit, the interaction energies
of the $AA$ and $AB$ subsystems are parametrized by the bound state energies $E_{AA}$ and $E_{AB}$.

We map $E_{AA}$ and $E_{AB}$ into the usual scattering lengths,
$a_{AA}$ and $a_{AB}$ through the relation $E\propto |a|^{-2}$.
Throughout most of this work we will focus on the
region close to unitarity in the $AB$ system, i.e. $|a_{AB}|\to \infty$ or $E_{AB}\to 0$.

We want now to introduce a new technique which can improve the numerical 
treatment of the problem, as already mentioned at the beginning of the 
section. Let us make a variable transformation in the set of couple integral 
equations (\ref{zre5a}), introducing 
\begin{equation}
\epsilon_3=R_y^2E_3~,~~\epsilon_{A\beta}=R_y^2E_{A\beta}~,~~\overline \mu=R_y\,\mu~, \label{e3r}
\end{equation}
with the momenta rescaled as:
\begin{equation}
\vec p_\perp\to R_y\,\vec p_\perp~,~~\vec q_\perp\to R_y\,\vec q_\perp~. \label{qpr}
\end{equation}
The transformation above corresponds to put  $R_y\to1$ in equations (\ref{zre5a}) and (\ref{rtauab}) 
provided the energies are substituted by (\ref{e3r}).

Introducing the following functional, 
\begin{equation}
{\cal F}(p_y)=\sum_{m}\delta\left(p_y-m\right) \ , 
\label{func}
\end{equation}
we can rewrite the set of coupled equations (\ref{zre5a}) as:
\begin{eqnarray}
\nonumber
&&f_{AB}(\vec q) =-\overline\tau_{AB}\left(E_3-\frac{{\cal A}+2}{2({\cal A}+1)} {\vec q}\,^2\right)
\int d^3p \, {\cal F}(p_y)\,
\left\{ K^{(1)\overline\mu}(\vec p,\vec q;\epsilon_3)\, f_{AA}(\vec p)
+  K^{(2)\overline\mu}(\vec q,\vec p;\epsilon_3)
 \,f_{AB}(\vec p)\right\} \\  
&&f_{AA}(\vec q) =-2\,\overline \tau_{AA}\left(\epsilon_3-\frac{{\cal A}+2}{4{\cal A}} {\vec q}\,^2\right) 
\int d^3p \, {\cal F}(p_y)\, K^{(1)\overline\mu}(\vec q,\vec p;\epsilon_3)
\,f_{AB}(\vec p)\, , 
   \label{zre5a1}
\end{eqnarray}
where we have identified $q_y\equiv n$  in the equation set (\ref{zre5a}). The kernels are defined by:
\begin{equation}
K^{(i)\overline\mu}(\vec q,\vec p;\epsilon_3)=
\left[\overline G^{(i)}_{0}(\vec q,\vec p;\epsilon_3)-\overline G^{(i)}_{0}(\vec q,\vec p;-\overline\mu^2)\right] 
\end{equation}
and the resolvents by
\begin{eqnarray}
\overline G^{(i)}_{0}(\vec q,\vec p;\epsilon)=\left[
\epsilon-\frac{{\cal A}+1}{2{\cal A}}\vec q\,^2 -\frac{{\cal A}+1}{{\cal A}+{\cal A}^{i-1}}
\vec p\, ^2-\frac{\vec q\cdot \vec p}{{\cal A}^{i-1}}
\right]^{-1}.
\end{eqnarray}
The two-body amplitudes for the new variables are given by
\begin{eqnarray}
\overline \tau_{A\beta;R}(\epsilon)=\left[
4\pi\,m_{A\beta}\ln\left({\sinh\pi\sqrt{-2\,m_{A\beta}\,\epsilon}\over\sinh\pi\sqrt{-2\,m_{A\beta}\,
\epsilon_{A\beta}}}\right)\right]^{-1}.
\end{eqnarray}

Let us proceed with the angular decomposition of the spectator functions:
\begin{equation}
f_{A\beta}(\vec q)=\sum_{L\,M}F^{A\beta}_{LM}(q^2)\,Y_{LM}(\theta_q,\phi_q)
\end{equation}
and the kernel:
\begin{equation}
K^{(i)\overline\mu}(\vec q,\vec p;\epsilon_3)=\sum_{ \overline L\,\overline M}K^{(i)\overline\mu}_{\overline L}(q, p;\epsilon_3)\,
Y^*_{\overline L \, \overline M}(\theta_q,\phi_q)
Y_{\overline L \, \overline M}(\theta_p,\phi_p).
\end{equation}
The angular momentum projection of the kernel is given by:
\begin{equation}
K^{(i)\overline\mu}_{L}(q, p;\epsilon_3)\,= 2\pi\int^{1}_{-1} d\cos\theta \, K^{(i)\overline\mu}(\vec q,\vec p;\epsilon_3)\,
P_L(\cos\theta).
\end{equation}

Performing the angular decomposition of Eq. (\ref{zre5a1}), where we used  the orthormalization 
of the spherical harmonics, we get the final form of the coupled integral equations for the bound-state 
of mass imbalanced systems for the $3D\to2D$ transition:
\begin{eqnarray}   
\nonumber
&&F_L^{AB}( q) =-\overline\tau_{AB}\left(E_3-\frac{{\cal A}+2}{2({\cal A}+1)} { q}\,^2\right) 
\sum_{L'}\int_0^\infty dp \, p^2 \, {\cal A}_{L,L'}(p^2)\, \left\{ K^{(1)\overline\mu}_{L}( p,q;\epsilon_3)
\, F^{AA}_{L'}(p) 
+  \, K^{(2)\overline\mu}_{L}( q,p;\epsilon_3)
\, F^{AB}_{L'}(p)\right\}
\\ 
&&F^{AA}_{L}(q) =-2\,\overline \tau_{AA}\left(\epsilon_3-\frac{{\cal A}+2}{4{\cal A}} {q}\,^2\right) 
 \sum_{L'}\int_0^\infty dp \, p^2 \, {\cal A}_{L,L'}(p^2)\, K^{(1)\overline\mu}_{L}( q,p;\epsilon_3)
\, F^{AB}_{L'}(p)    \, ,
\end{eqnarray}
where we have dropped the reference to the magnetic quantum number due to the cylindrical symmetry of the 
squeezed set up for the $AAB$ system. The matrix element of the functional (\ref{func}) for angular momentum states is
\begin{equation}
{\cal A}_{L,L'}(p^2)=\frac{1}{2p}\sqrt{(2L+1)(2L'+1)} \sum_m\theta\left(1-\frac{|m|}{p}\right) P_L\left(\frac{m}{p}\right) P_{L'}\left(\frac{m}{p}\right) \  .
\end{equation}

\subsection{Transition 2D$\rightarrow$1D}

The procedure here is very close to the one applied in the previous section. We start now 
with one less dimension. The coupled integral equations for the spectator functions, $f_{AA}$ and 
$f_{AB}$ reads:
\begin{eqnarray}
\nonumber
&&f_{AB}(\tilde q) =-\tau_{AB;R_x}\left(E_3-\frac{{\cal A}+2}{2({\cal A}+1)}  \tilde q^2 \right) 
\sum_m\int \frac{dp}{R_x} \left\{G^{(1)}_{0R_x}(\tilde p,\tilde q;E_3)\, f_{AA}(\tilde p)
+ G^{(2)}_{0R_x}(\tilde q,\tilde p;E_3)\, f_{AB}(\tilde p)\right\},\\  
&&f_{AA}(\tilde q) =-2\,\tau_{AA;R_x}\left(E_3-\frac{{\cal A}+2}{4{\cal A}} \tilde q^2\right)
\sum_m\int \frac{dp}{R_x}G^{(1)}_{0R_x}(\tilde q,\tilde p;E_3)\, f_{AB}(\tilde p)
\label{zre5a1d}
\end{eqnarray}
where  $\tilde q\equiv (q,n)$, $\tilde p\equiv (p,m)$ and 
\begin{eqnarray}
\tilde q^2=q^2+{n^2\over R_x^2}\, ,\,\, \tilde p^2=p^2+{m^2\over R_x^2} \,\, {\rm and} \,\,\,
\tilde q\cdot \tilde p= q\,p +\frac{n\;m}{R_x^2}
\end{eqnarray}
The resolvents are defined by:
\begin{equation}
\left[G^{(1)}_{0R_x}(\tilde q,\tilde p;E)\right]^{-1}=
E-\tilde p^2-\tilde q\cdot \tilde p-\frac{{\cal A}+1}{2{\cal A}}\tilde q^2 
\,\,\, , \,\,\,
\left[G^{(2)}_{0R_x}(\tilde q,\tilde p;E)\right]^{-1}= E
-\frac{\tilde q\cdot \tilde p}{{\cal A}}-\frac{{\cal A}+1}{2{\cal A}}(\tilde q^2+\tilde p^2)
\label{greenq1d}\end{equation}
The two-body amplitudes for finite $R_x$ are given by
\begin{eqnarray}
\nonumber
&&\tau^{-1}_{A\beta;R_x}(E)
 =\frac2R_x\,m_{A\beta}\,\left\{\sum_n\int  
\frac{dp}{\tilde E-
p^2-\frac{n^2}{R_x^2}} 
-\sum_n\int  
\frac{dp}{\tilde E_{A\beta}-
p^2-\frac{n^2}{R_x^2}} \right\} 
\ , \label{rtauab1d}
\end{eqnarray}
with $\beta\equiv A$ or $ B$, $\tilde E=2\,m_{A\beta} E$ ($E<0$)
and $\tilde E_{A\beta}=2\,m_{A\beta} E_{A\beta}$ and we chose the bound-state pole at $E_{A\beta}$ 
for each $R_x$. The reduced mass is $m_{A\beta}=m_A\, m_\beta/(m_A+ m_\beta)$.
The interaction energies of the $AA$ and $AB$ subsystems are parametrized by the bound state 
energies $E_{AA}$ and $E_{AB}$.

Here, we continue to follow the procedure of the last subsection. Consider the variable 
transformation as follows
\begin{equation}
\epsilon_3=R_x^2E_3~,~~\epsilon_{A\beta}=R_x^2E_{A\beta}~, \label{e3r1d}
\end{equation}
with the momentum rescaled as:
\begin{equation}
p\to R_x\, p~,~~q\to R_x\, q~. \label{qpr1d}
\end{equation}
The transformation above corresponds to put  $R_x\to1$ in equations (\ref{zre5a1d}) 
and (\ref{rtauab1d}) provided the energies are substituted by (\ref{e3r1d}).

Introducing the functional given by Eq. (\ref{func}), we can rewrite the set of coupled 
equations (\ref{zre5a1d}) as:
\begin{eqnarray}
\nonumber
&&f_{AB}(\vec q) =-\overline\tau_{AB}\left(E_3-\frac{{\cal A}+2}{2({\cal A}+1)} {\vec q}\,^2\right)  
\int d^2p \, {\cal F}(p_x)\,
\left\{ K^{(1)}(\vec p,\vec q;\epsilon_3)\, f_{AA}(\vec p)
+K^{(2)}(\vec q,\vec p;\epsilon_3)
\,f_{AB}(\vec p)\right\}
\\ 
&&f_{AA}(\vec q) =-2\,\overline \tau_{AA}\left(\epsilon_3-\frac{{\cal A}+2}{4{\cal A}} {\vec q}\,^2\right) 
\int d^2p \, {\cal F}(p_x)\, K^{(1)}(\vec q,\vec p;\epsilon_3)
\,f_{AB}(\vec p),
\label{zre5a11d}
\end{eqnarray}
where we have identified $q_x\equiv n$ in the equation set (\ref{zre5a1d}). The kernels are defined by:
\begin{eqnarray}
K^{(i)}_{0}(\vec q,\vec p;\epsilon)=\left[
\epsilon-\frac{{\cal A}+1}{2{\cal A}}\vec q\,^2 -\frac{{\cal A}+1}{{\cal A}+
{\cal A}^{i-1}}\vec p\, ^2-\frac{\vec q\cdot \vec p}{{\cal A}^{i-1}}
\right]^{-1}
\end{eqnarray}
The two-body amplitudes for the new variables are given by
\begin{equation}
\overline \tau^{-1}_{A\beta;R}(\epsilon)=
2\pi \,m_{A\beta}
\sum^\infty_{n=-\infty}\left[
\frac{1}{\sqrt{ -m_{A\beta}\,\epsilon_{A\beta}+n^2}} 
-\frac{1}{\sqrt{-m_{A\beta}\, \epsilon+n^2} }\right]
\ , \label{tauq1d}
\end{equation}

The angular decomposition of the spectator functions is given by:
\begin{equation}
f_{A\beta}(\vec q)=\frac{1}{\sqrt{2\pi}}
\sum_{M}F^{A\beta}_{M}(q^2)\,\exp\left(i\, M\,\phi_q\right)
\end{equation}
and the kernel:
\begin{equation}
K^{(i)}(\vec q,\vec p;\epsilon_3)=\frac{1}{2\pi}\sum_{ \overline M}K^{(i)}_{\overline M}(q, p;\epsilon_3)\,\exp\left(i \,\overline M (\phi_q-\phi_p)\right)
\end{equation}
The angular momentum projection of the kernel is given by:
\begin{equation}
K^{(i)}_{M}(q, p;\epsilon_3)\,= \int^{2\pi}_{0} d\phi \, K^{(i)}(\vec q,\vec p;\epsilon_3)\,\exp(-i \,M\,\phi)
\end{equation}

Performing the angular decomposition of Eq. (\ref{zre5a11d}) and using the orthonormalization of the 
angular states we have the final form of the coupled integral equations for the bound state 
of mass imbalanced systems for the $2D\to1D$ transition:
\begin{eqnarray}
\nonumber
&&F_M^{AB}( q) =-\overline\tau_{AB}\left(E_3-\frac{{\cal A}+2}{2({\cal A}+1)} { q}
\,^2\right)
\sum_{M'}\int_0^\infty dp \, p \, {\cal B}_{M-M'}(p^2)\, 
\left\{ K^{(1)}_{M}( p,q;\epsilon_3)
\, F^{AA}_{M'}(p)
+\, K^{(2)}_{M}( q,p;\epsilon_3)
\, F^{AB}_{M'}(p)\right\} \ ,
\\
&&F^{AA}_{M}(q) =-2\,\overline \tau_{AA}\left(\epsilon_3-\frac{{\cal A}+2}{4{\cal A}} {q}\,^2\right) 
\sum_{M'}\int_0^\infty dp \, p \, {\cal B}_{M-M'}(p^2)\, K^{(1)}_{M}( q,p;\epsilon_3)
\, F^{AB}_{M'}(p)\, ,
\end{eqnarray}
where the matrix elements of the functional (\ref{func}) in the 2D angular momentum states are:
\begin{equation}
{\cal B}_{M}(p^2)=
\frac1p\sum_m \Theta\left(1-\frac{|m|}{p} \right) 
\frac{ \cos\left(M\,\cos^{-1}(m/p) \right)}{\sqrt{1-(m/p)^2}}
 \, . \label{aa21d}
\end{equation}

\subsection{Physical interpretation of the compactification procedure}
Our technique, with an appropriate association between the compactification radius and harmonic oscillator length, as already discussed, 
exhibit the same behavior of the binding energy when the two-body system is squeezed 
from 3D $\to$ 2D and 2D $\to$ 1D. It is only the limiting values at integer dimensions that depend on the potential details. 
Therefore,
the input, namely the two-body amplitudes Eq.~\eqref{tauq3d} and Eq.~\eqref{tauq1d}, entering the kernel of the coupled momentum space Faddeev
equations express the squeezing quantitatively. 

The other important quantity that enters is the three-body Green's functions, 
Eq.~\eqref{greenq2d} and Eq.~\eqref{greenq1d} with quantized momentum. These are the other components of the kernel of 
the bound state integral equations that drive the trimer from 3D $\to$ 2D and 2D $\to$ 1D, respectively. The 
compactification technique introduces the quantization of the relative momentum of the spectator particle with 
respect to the center of mass of the other two. 
At this point it is useful to recall that the Green's functions represent the 
one-particle exchange mechanism, which produces the Efimov long-range potential and also contains a Yukawa
potential when $m_B<<m_A$ due to the effective interaction between the heavy particle A and the light particle B in the pair with the third 
particle A \cite{fonseca}, schematically represented by A + (AB) $\to$ (AB) + A. 
Furthermore, in the present three-body model the
spectator function is analogous to a relative two-body wave function (an old interpretation given by Mitra \cite{mitra} when formulating
the integral equations for the bound and scattering states for one-term separable potentials). 
In light of these previous developments, 
the present three-body
model has dynamics that can be interpreted as an effective two-body dynamics. This implies that the compactification method, which works quantitatively on the two-body level as we have shown, preserves the physical picture and thus should also work both qualitatively and to a high degree also quantitatively at the three-body level.


\begin{thebibliography}{99}

\bibitem{greene-review2017} C.~H. Greene, P. Giannakeas, and J. Perez-Rios, arXiv:1704.02029.
\bibitem{naidon-review2017} P. Naidon and S. Endo, Rep. Prog. Phys. {\bf 80}, 056001 (2017).
\bibitem{zinner2014} N.~T. Zinner, Few-Body Syst. {\bf 55}, 599 (2014).
\bibitem{dincao2017} J.~P. D'Incao, arXiv:1705.10860.

\bibitem{efi70} V. Efimov, Yad. Fiz \textbf{12}, 1080 (1970); Sov. J. Nucl. Phys. \textbf{12}, 589 (1971).

\bibitem{kraemer2006} T. Kraemer {\it et al.}, Nature {\bf 440}, 315 (2006).
\bibitem{dorner2015} M. Kunitski {\it et al.}, Science {\bf 348}, 551 (2015).
\bibitem{chin2010} C. Chin, R. Grimm, P.~S. Julienne, and E. Tiesinga, Rev. Mod. Phys. {\bf 82}, 1225 (2010).
\bibitem{bloch2008} I. Bloch, J. Dalibard, and W. Zwerger, Rev. Mod. Phys. {\bf 80}, 885 (2008).
\bibitem{deng2016} S. Deng {\it et al.}, Science {\bf 353}, 371 (2016).
\bibitem{gross2009} N. Gross {\it et al.}, Phys. Rev. Lett. {\bf 103}, 163202 (2009).
\bibitem{knoop2009} S. Knoop {\it et al.}, Nature Phys. {\bf 5}, 227 (2009).
\bibitem{zaccanti2009} M. Zaccanti {\it et al.}, Nature Phys. {\bf 5}, 586 (2009).
\bibitem{williams2009} J.~R. Williams {\it et al.}, Phys. Rev. Lett. {\bf 103}, 130404 (2009).
\bibitem{gross2010} N. Gross {\it et al.}, Phys. Rev. Lett. {\bf 105}, 103203 (2010).
\bibitem{lompe2010} T. Lompe {\it et al.}, Science {\bf 330}, 940 (2010).
\bibitem{nakajima2010} S. Nakajima {\it et al.}, Phys. Rev. Lett. {\bf 105}, 023201 (2010).
\bibitem{berninger2011} M. Berninger {\it et al.}, Phys. Rev. Lett. {\bf 107}, 120401 (2011).
\bibitem{machtey2012} O. Machtey {\it et al.}, Phys. Rev. Lett. {\bf 108}, 210406 (2012).
\bibitem{wild2012} R.~J. Wild {\it et al.}, Phys. Rev. Lett. {\bf 108}, 145305 (2012).
\bibitem{knoop2012} S. Knoop {\it et al.}, Phys. Rev. A {\bf 86}, 062705 (2012).
\bibitem{roy2013} S. Roy {\it et al.}, Phys. Rev. Lett. {\bf 111}, 053202 (2013).
\bibitem{dyke2013} P. Dyke, S.~E. Pollack, and R.~G. Hulet, Phys. Rev. A {\bf 88}, 023625 (2013).
\bibitem{huang2014} B. Huang {\it et al.}, Phys. Rev. Lett. {\bf 112}, 190401 (2014).

\bibitem{bruch1979} L.~W. Bruch and J.~A. Tjon, Phys. Rev. A {\bf 19}, 425 (1979). 
\bibitem{nielsen1997} E. Nielsen, D.~V. Fedorov, and A.~S. Jensen, Phys. Rev. A {\bf 56}, 3287 (1997).
\bibitem{brodsky2006} I.~V. Brodsky {\it et al.}, Phys. Rev. A {\bf 73}, 032724 (2006).
\bibitem{kartavtsev2006} O.~I. Kartavtsev and A.~V. Malykh, Phys. Rev. A {\bf 74}, 042506 (2006).
\bibitem{pricoupenko2010} L. Pricoupenko and P. Pedri, Phys. Rev. A {\bf 82}, 033625 (2010).
\bibitem{helfrich2011} K. Helfrich and H.-W. Hammer, Phys. Rev. A {\bf 83}, 052703 (2011).
\bibitem{volosniev2013} A.~G. Volosniev {\it et al.}, Eur. Phys. J. D {\bf 67}, 95 (2013).

\bibitem{nishida2013} Y. Nishida, S. Moroz, and D.~T. Son, Phys. Rev. Lett. {\bf 110}, 235301 (2013).
\bibitem{volosniev2014} A.~G. Volosniev {\it et al.}, J. Phys. B:At.Mol.Opt.Phys. {\bf 47}, 185302 (2014).
\bibitem{gao2015} C. Gao, J. Wang, and Z. Yu, Phys. Rev. A {\bf 92}, 020504 (2015).
\bibitem{efremov2014} M.~A. Efremov and W.~P. Schleich, arXiv:1407.3352.

\bibitem{nie01} E. Nielsen, D.~V. Fedorov, A.~S. Jensen, and E. Garrido,
  Phys. Rep. \textbf{347}, 373 (2001).


\bibitem{schroder1995} M.~E. Peskin and D.~V. Schroeder: \emph{An Introduction to Quantum Field Theory} (Avalon Publishing, 1995).
\bibitem{valiente2012} M. Valiente {\it et al.}, Phys. Rev. A {\bf 86}, 043616 (2012).



  

\bibitem{lev14} J. Levinsen, P. Massignan, and M.~M. Parish,
Phys.Rev. X {\bf 4} , 031020 (2014).

\bibitem{yam15} M. T. Yamashita, F.~F. Bellotti, T. Frederico,
  D.~V. Fedorov, A.~S. Jensen, N.~T. Zinner, J. Phys. B:At. Mol. Opt. Phys. {\bf 48}, 025302 (2015).

\bibitem{levinsen2009} J. Levinsen, T.~G. Tiecke, J.~T.~M. Walraven, and D.~S. Petrov, Phys. Rev. Lett. {\bf 103}, 153202 (2009).
\bibitem{nishida2008} Y. Nishida and S. Tan, Phys. Rev. Lett. {\bf 101}, 170401 (2008).

\bibitem{barontini2009} G. Barontini {\it et al.}, Phys. Rev. Lett. {\bf 103}, 043201 (2009).
\bibitem{bloom2013} R.~S. Bloom {\it et al.}, Phys. Rev. Lett. {\bf 111}, 105301 (2013).
\bibitem{pires2014} R. Pires {\it et al.}, Phys. Rev. Lett. {\bf 112}, 250404 (2014).
\bibitem{tung2014} S.~K. Tung {\it et al.}, Phys. Rev. Lett. {\bf 113}, 240402 (2014).
\bibitem{maier2015} R.~A.~W. Maier {\it et al.}, Phys. Rev. Lett. {\bf 115}, 043201 (2015).
\bibitem{ulmanis2016} J. Ulmanis {\it et al.}, Phys. Rev. Lett. {\bf 117}, 153201 (2016). 
\bibitem{wacker2016} L.~J. Wacker {\it et al.}, Phys. Rev. Lett. {\bf 117}, 163201 (2016).
\bibitem{johansen2016} J. Johansen {\it et al.}, arXiv:1612.05169.

\bibitem{supmat} See Supplementary Materials.


\bibitem{fre11} T. Frederico, L. Tomio, A. Delfino, M.~R. Hadizadeh and M.~T. Yamashita, Few-Body Systems {\bf 51} 87 (2011).

\bibitem{adh95a} S.~K. Adhikari, T. Frederico, I.~D. Goldman, Phys. Rev.  Lett. {\bf 74}, 487 (1995).

\bibitem{adh95b} S.~K. Adhikari, T. Frederico, Phys. Rev. Lett. {\bf 74}, 4572 (1995).

\bibitem{mitroy2013} J. Mitroy {\it et al.}, Rev. Mod. Phys. {\bf 85}, 693 (2013).

\bibitem{bel11} F.~F. Bellotti {\it et al.}, J. Phys. B:At. Mol. Opt. Phys. {\bf 44}, 205302 (2011).
\bibitem{bel12} F.~F. Bellotti {\it et al.}, Phys. Rev A {\bf 85}, 025601 (2012).
\bibitem{bel13b} F.~F. Bellotti {\it et al.}, J. Phys. B:At. Mol. Opt. Phys. {\bf 46}, 055301 (2013).

\bibitem{note-on-dim} The dimensional requirement for the Efimov effect to occur, $2.3<d<3.8$ \cite{nie01}, 
depends generally on the masses in the system and the numbers will thus change for our ratio 
of $m_B/m_A=6/133$, although the expected modification is small, see
D.~S. Rosa, T. Frederico, G. Krein, and M.~T. Yamashita, arXiv:1707.06616.

\bibitem{jen03} A.~S. Jensen and D.~V. Fedorov, Europhys.Lett. {\bf 62},
  336 (2003).
\bibitem{yam13}  M.~T. Yamashita, F.~F. Bellotti, T. Frederico, D.~V. Fedorov,
A.~S. Jensen, N.~T. Zinner, Phys. Rev. A {\bf 87}, 062702 (2013).

\bibitem{zeldovich1960} Y.~B. Zel'dovich, Sov. J. Solid State {\bf 1}, 1497 (1960).
\bibitem{jensen2004} A.~S. Jensen, K. Riisager, D.~V. Fedorov, and E. Garrido, Rev. Mod. Phys. {\bf 76}, 215 (2004).
\bibitem{braaten2006} E. Braaten and H.-W. Hammer, Phys. Rep. {\bf 428}, 259 (2006).

\bibitem{portegies2011} J. Portegies and S.~Kokkelmans, Few-Body Syst. {\bf 51}, 219 (2011).

\bibitem{san16} J.~H. Sandoval, F.~F. Bellotti, A.~S. Jensen, M.~T. Yamashita,
Phys. Rev. A {\bf 94}, 022514 (2016).




\bibitem{ziegelmann} E.~W. Schmid and H. Ziegelmann: \emph{The Quantum Mechanical Three-Body Problem}, 
(Pergamon Press 1974).

\bibitem{mitroy2013} J. Mitroy {\it et al.}, Rev. Mod. Phys. {\bf 85}, 693 (2013).

\bibitem{suzuki} Y. Suzuki and K. Varga: \emph{Stochastic Variational Approach to 
Quantum-Mechanical Few-Body Problems}, (Springer, 1998).

\bibitem{adhikari} S.~K. Adhikari, T. Frederico, I.~D. Goldman, Phys. Rev. Lett. {\bf 74}, 487 (1995); 
S.~K. Adhikari, T. Frederico, Phys. Rev. Lett. {\bf 74}, 4572 (1995).


\bibitem{fonseca} A.~C. Fonseca, E.~F.  Redish,  P.~E. Shanley, Nucl. Phys. A {\bf 320}, 273 (1979)

\bibitem{mitra} A.~N. Mitra, Adv. Nucl. Phys. {\bf 3}, 1 (1969). 


  
\end{thebibliography}
\end{document}